\documentclass[letter]{aa} 
\usepackage[normalem]{ulem} 
\usepackage{graphicx}
\usepackage{txfonts}
\begin{document}

   \title{Cosmic reionization by primordial cosmic rays}

   \author{M. Tueros\inst{1}, M.V. del Valle\inst{1}, 
          \and G.E. Romero\inst{1}$^{,}$\inst{2}
    }
\institute{Instituto Argentino de Radioastronom\'{\i}a, C.C.5, (1894) Villa Elisa, Buenos Aires, Argentina\\
              \email{mjtueros@iar-conicet.gov.ar}
         \and Facultad de Ciencias Astron\'omicas y Geof\'{\i}sicas, Universidad Nacional de La Plata, Paseo del Bosque, 1900 La Plata, Argentina\\
             }

   \date{Received; accepted}

 
 \abstract
{After the so-called cosmic recombination, the expanding universe entered into a period of darkness since most of the matter was in a neutral state. About a billion years later, however, the intergalactic space was once again ionized. The process, known as the cosmic reionization, required the operation of mechanisms that are not well understood.  Among other ionizing sources, Population III stars, mini-quasars, and X-ray emitting microquasars have been invoked.}
   {We propose that primordial cosmic rays, accelerated at the termination points of  the jets of the first microquasars, may have contributed to the reionization of the intergalactic space as well.}
   {We quantify  the ionization power of cosmic rays (electrons and protons) in the primordial intergalactic medium. This power is calculated using extensive particle cascade simulations.}
   {We establish that, depending on the fraction of electrons to protons accelerated in the microquasar jets, cosmic rays should have contributed to the reionization of the primordial intergalactic medium as much as X-rays from  microquasar accretion disks. If the primordial magnetic field was of the order of $10^{-17}$ G, as some models suggest, cosmic rays had an important role in ionizing the neutral material far beyond the birth places of the first stars.}
{}


\authorrunning{M.Tueros et al.}
\titlerunning{Reionization of the universe}

   \maketitle
%

\section{Introduction}\label{intro}

The standard Big Bang model of the expanding universe requires that when the temperature of the primordial plasma dropped below the combination temperature, around $380 000$ yr after the beginning of the current expansion phase, the combination of protons and electrons formed neutral gas, allowing the radiation to decouple with matter. The universe then entered in to a ``dark age", which lasted up to about a billion years (e.g., Ellis, Maartens \& MacCallum 2012). How the universe was reionized is a major topic in current cosmology. The influence of cosmic rays (CRs) from the first galaxies on the reionization process has been studied by Biermann \& Nath (1993), who showed that models with strong evolution and early ($z\sim10$) galaxy formation could account for the reionization of the universe. Later measurements of the Wilkinson Microwave Anisotropy Probe (WMAP) satellite (Kogut et al. 2003) hinted that the reionization could have occurred as early as $11< z <30$, requiring much older ionization sources.

The decay of exotic primordial particles is sometimes invoked as a possible ionization source (i.e., Belikov \& Hooper 2009, Natarajan \& Schwarz 2010, Iocco 2010). A more conservative hypothesis, requiring no new physics, is that the reionization was produced or at least initiated by the first stellar objects.

The formation of the first stars of zero metallicity, at redshift $z\sim20$, resulted in the injection of a large number of ultraviolet (UV) photons (e.g., Loeb 2010). However, it seems difficult for these photons to interact with neutral gas at large distances from the stars, given the high-column densities of primordial {star-forming} clouds. Recently, Mirabel et al. (2011) have proposed that X-rays from accreting black holes in early binary systems might have played a crucial role, because of the longer mean free path of X-rays with respect to UV radiation.
 
The first generations of microquasars (MQs) should have not only produced copious X-rays, but also relativistic particles through their jets. 

The jets can propagate hundreds of parsecs and escape the original cloud where the star formation took place. Once the jets were in what would become the intergalactic medium (IGM), the termination shocks could reaccelerate protons and electrons up to relativistic energies. Then, these particles would diffuse, ionizing the medium they encountered. In this letter we offer a quantitative estimate of the ionizing power of these particles as they diffused through the early universe, and we compare this power with that of the X-ray emission of the same MQs 

In the next section, we describe the simulations of CR propagation in the early universe that we have performed to estimate the average ionizing power per primordial CR. Then, in Section 3, we develop the microquasar scenario and ponder the relative strength of both the X-ray and cosmic-ray output of these systems. We close with a discussion of our results in Section 4.


\section{Ionization power of cosmic rays}\label{power}


We have used a heavily modified version of the AIRES code (see AIRES Manual) to estimate the ionization power of electrons and protons injected directly into the IGM by the jets from microquasars. Detailed simulations of the ionization power of electrons in the IGM have been reported by Valdes et al. (2010) for fixed values of $z$ and energies up to the TeV range. For this letter, a new simulation that takes into account the evolution of the IGM conditions was necessary, as high energy particles can survive through al the reionization period.

We have modified AIRES propagation routines to include a redshift-dependent monochromatic photon field to simulate the cosmic microwave background (CMB) and a material medium of hydrogen atoms with redshift-dependent number density. We added inverse Compton scattering and $e^{\pm}$ photo-pair production for electrons, positrons, and photons. Neutron decays have been also included. All relevant hadronic interactions are present in the code, including secondary meson interactions. 

When a particle cascade develops, most of the ionization in the traversed medium is produced by low-energy particles, especially electrons and photons in and below the keV energy range. Unfortunately, the full simulation of all processes leading to particles in and below the keV range would require humongous amounts of CPU time. To circumvent this problem, the generation of particles below a certain threshold (100 keV for electrons and photons, and 500 keV for other particles) has not been directly simulated. Instead, the generation of low-energy particles is represented by an averaged energy loss per amount of traversed matter, which has been subtracted from all charged particles during their propagation.

As low-energy particles lose most of their energy through ionization, the subtracted energy ($E_{\rm lost}$) has been considered to be ultimately deposited in the traversed medium through ionization. The number of ions that would have been generated by these low-energy particles can then be estimated using the mean energy loss per ionization event $I_{\rm H}$. We take $I_{\rm H}$ to be $\approx 36$~eV, considering that $10.2$~eV {goes} to ionize the atom and $\approx 25.8$~eV corresponds to the average kinetic energy of the outgoing electron, of which $22.3$~eV are lost in excitations of atomic levels and $3.4$~eV in heating of the gas (Spitzer \& Tomasko 1968).

To further speed up the simulations, we have discarded particles that fell below the low-energy threshold during their propagation. The energy carried away by these particles ($E_{\rm dis}$) has also been considered to be ultimately deposited in the medium through ionization. Then, the total number of produced ions is:
\begin{equation}
N_{\rm ions}=\frac{(E_{\rm lost}+E_{\rm dis})}{I_{\rm H}}. \label{eq:ionization}
\end{equation}
The IGM has a redshift dependence. We have adopted a density of the primordial IGM of $ n_{\rm H} = 2.5 \times 10^{-30}(1+z)^3$~g~cm$^{-3}$ (e.g., Ellis et al. 2012). The CMB has been  considered monoenergetic with photon energy $E_{\rm CMB} = 3.75 \times 10^{-4}(1+z)$~eV, and a photon density  $u_{\rm CMB} = 0.05\,(1+z)^3$ cm$^{-3}$. For the magnetic field of the IGM, we {have taken} the value $B = 10^{-17}$~G (e.g., Stacy \& Bromm 2007; Loeb 2010;  Bromm 2013). Note that several theories proposing the generation of magnetic fields during the inflation phase yield magnetic fields ten orders of magnitude higher at $z=$20 (Widrow et al. 2012). The magnetic field strength has almost no effect on the ionization efficiency of CRs, but it will strongly affect their diffusion scale.

The model {we have} used for the evolution of the universe is that of the standard spatially-flat six-parameter $\Lambda$CDM cosmology, with a Hubble constant $H_{0}=67.3 \pm 1.2$ km~s$^{-1}$~Mpc$^{-1}$ and a matter density parameter $\Omega_{\rm m}=0.315 \pm 0.017$, in accordance with latest results from the Planck collaboration (2014).

Under these assumptions, simulations of electrons in the 1 MeV - 100 TeV and protons in the 1 GeV - 100 TeV energy range have been generated. Particles start at redshift $z = 19$ and they are propagated through the IGM until they reach $z = 5$, the epoch at which the reionization is considered to be complete (Loeb 2010). During this span, particles can traverse up to $\sim$ 3.1~g~cm$^{-2}$ of matter.

\begin{figure} 
\begin{center}
\includegraphics[trim=0cm 0cm 0cm 0cm, clip=true,width=.5\textwidth,angle=0]{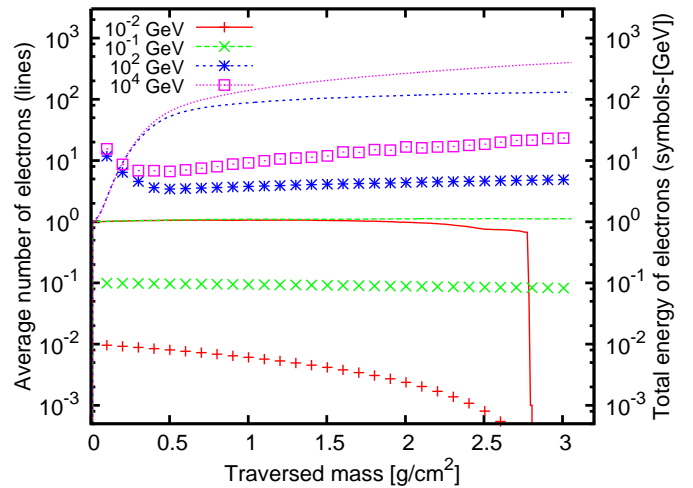}
\caption{{Average longitudinal development of electrons of several energies injected in the IGM at redshift $z = 19$}}
\label{gr:longitudinal}
\end{center}
\end{figure}

Our simulations for electron primaries show that below the energy threshold for interactions with the CMB, roughly 1 GeV, no particle cascades are produced. An example of the average longitudinal development of 10 MeV electrons is shown in Figure \ref{gr:longitudinal}. At these energies, electrons propagate losing their energy mainly by ionization of the occasional hydrogen atoms they encounter on their path.

Electrons below 10 MeV lose all their energy in the IGM before the reionization epoch ends, producing a number of ionizations (i.e., an ionization power) proportional to their initial energy. 

Above 10 MeV, electrons survive the reionization epoch, as can be seen on the average longitudinal profile for 100 MeV electrons shown in Fig. \ref{gr:longitudinal}. This gives a plateau in the ionization power for electrons between 10 MeV and 1 GeV. The energy deposited through ionization by a single particle for a fixed amount of traversed matter has little dependency on its energy.

\begin{figure} 
\begin{center}
\includegraphics[trim=0cm 0cm 0cm 0cm, clip=true,width=.5\textwidth,angle=0]{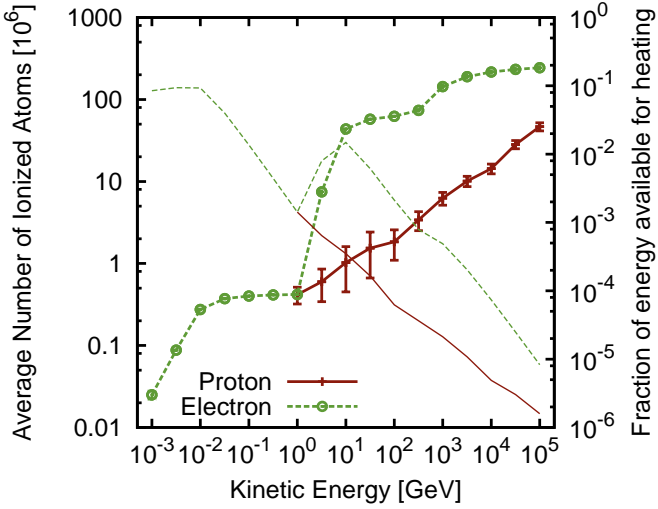}
\caption{Ionization power (left axis, points with lines) and fraction of the primary energy available for gas heating (right axis, lines only) of particles injected in the IGM at redshift $z = 19$ as a function of the primary energy.}
\label{gr:ionizationrange}
\end{center}
\end{figure}

Once the energy of the primary electrons reaches the threshold of inverse Compton with the CMB, particle cascades start being generated. The number of particles quickly rises, as does the total energy lost through ionization. At some point these particles start having enough energy to survive the reionization epoch. This yields the second plateau in Fig. \ref{gr:ionizationrange}. 

A second sudden, but smaller rise in the ionizing power is produced when secondaries also reach the CMB interaction threshold, triggering more cascades. An example of this type of cascades is shown in Fig. \ref{gr:longitudinal} for 10 TeV electrons, where it can be seen that the number of electrons and the energy they carry continues to rise, in contrast with the 100 GeV case.

For proton primaries, we started our calculations at 1 GeV. At these energies, protons have a large variability in their ionization power as the 3.1~g~cm$^{-2}$ of traversed matter is well below the mean free path of $p-p$ interactions. Protons that do not interact survive the reionization epoch and deposit very little energy, yielding a small ionization power. Protons that do interact generate pions that promptly decay into muons and photons that in turn generate cascades that do have a large ionization power. This yields a large variability in the ionization power of protons, especially at low energies, as can be seen in Fig. \ref{gr:ionizationrange}. We show in this figure also the fraction of energy of the primary particles available for heating the medium. It can be seen that particles below 10 MeV have a heating efficiency of 0.1 and could have a sensible contribution not only to ionizing but also to heating the IGM. A detailed discussion of heating of the primordial IGM by CRs will be presented elsewhere.

In the next section, we will use these results to calculate the total ionization power of a microquasar that injects particles directly into the IGM with a given injection spectrum of the form $N(E) = N_{0} E^{-\beta}$, where  $N_{0}$ is the normalization constant. As the ionization power of particles I(E) is a function the energy, we characterize the ionization power of the particles with the equivalent ionization power $I_{E_{\rm c}}$, defined as the ionization power that particles of characteristic energy $E_{\rm c}$ would need to have if the energy budget of the microquasar jet goes entirely to particles of that energy:
\begin{eqnarray}
\label{eq:1GeVioniz}
     I_{E_{\rm c}} = \frac{N_{\rm ionizations}}{E_{\rm total}/{E_{\rm c}}}   &=& \frac{\int^{E_2}_{E_1}{ I(E)\,N_{0}\,E^{-\beta} dE}}{\left( \int^{E_2}_{E_1}{N_{0}\,E^{-\beta} dE}\right){E^{-1}_{\rm c}}}.\\ \nonumber   
\end{eqnarray}For a spectral index  $\beta=2.2$ (as expected from nonrelativistic shock acceleration, e.g., Drury 1983), and integration limits $E_{1}$=1~GeV for protons, $E_{1}$=1~MeV for electrons,  and  $E_{2}$=100~TeV, we obtain $I^{p}_{\rm{1\,GeV}} \approx 10^{5}$ ionizations per injected 1 GeV proton, and $I^{e}_{1\,\rm{MeV}} \approx 1.5 \times 10^{4}$ ionizations per injected 1 MeV electron.

 
 \section{Cosmic-ray ionization power from primordial microquasar jets}\label{jets}

As mentioned above, MQs produce collimated relativistic jets (Mirabel \& Rodr{\'\i}guez 1999). The supersonic impact of the relativistic fluid with the external medium produces a termination shock (Heinz \& Sunyaev 2002). At these shocks, relativistic particles (in principle both protons and electrons) are expected to be accelerated via a diffusive first-order mechanism such as Fermi I (e.g., Zealey, Dopita \& Malin 1980; Bordas et al. 2009). The relativistic particles cool in the acceleration region producing nonthermal radiation, and the maximum energy achieved is the result of the balance between energy gain and losses.  In these systems, particles are more efficiently accelerated at the reverse shock, which is the fastest. Most of the relativistic particles escape from the acceleration region, without losing much energy,  and diffuse into the ambient medium (see Bordas et al. 2009). Jets of galactic MQs are known to extend up to a distances of several parsecs from the accreting black hole, but some extragalactic microquasar jets can reach up to 300 pc from the binary system (e.g., Pakull, Soria \& Motch 2010; Middleton et al. 2013). Primordial MQs are expected to be very powerful: they should be formed by a very massive donor star with almost zero metallicity and a black hole of ten or more solar masses. Since Pop III stars have no winds (that are line-driven,{ e.g., Krti\v{c}ka \& Kub\'at 2006, 2009}), these early MQs must have been in a Roche-lobe overflow accretion regime. So they would be more similar to Galactic low-mass MQs, where jets carry a power that is a significant fraction of the Eddington accretion rate (Romero \& Vila 2008; Vila \& Romero 2010; Vila et al. 2012).  With powers reaching $\sim 10^{38-39}$ erg s$^{-1}$, the jets of these Pop III MQs might have been able to inject relativistic particles outside the regions of the primordial clouds where they formed. These clouds are thought to have radii $\sim$ 0.1-1 kpc (e.g., Stacy \& Bromm 2007).

The ratio of ionizations produced by CRs to those  produced by X-rays, $\chi$, can be estimated as:

\begin{equation}\label{chi}
\chi = \frac{N_{\rm io,CR}\,N_{\rm CR}}{N_{\rm io,X}\,N_{\rm X}}\frac{\epsilon_{\rm CR}}{\epsilon_{\rm X}}\,\kappa = \frac{(N_{{\rm io}, e}\,N_{e} + N_{{\rm io},p}\,N_{p})}{N_{\rm io,X}\,N_{\rm X}}\frac{\epsilon_{\rm CR}}{\epsilon_{\rm X}}\,\kappa ,
\end{equation}

where $N_{\rm io,CR}, N_{{\rm io},e}, N_{{\rm io},p}$, and $N_{\rm io,X}$ are the number of ionizations produced by a CR (electron/proton) and an X-ray photon, respectively; $N_{\rm CR}$ is the number of CRs and $N_{\rm X}$ is the number of X-ray photons.  The parameters $\epsilon_{\rm CR}$ and  $\epsilon_{\rm X}$ are the fraction of CRs or photons escaping into the IGM, respectively. $\kappa$ is the fraction of X-ray binaries that produce jets. The number of ionizations {$N_{p,e}$ and $N_{\rm X}$ can be estimated as}

\begin{equation}
N_{p, e} = \frac{L_{e, p}\,t_{\rm CR}} {\langle E_{e,p} \rangle},\,\,N_{\rm X} = \frac{L_{\rm X}\,t_{\rm X}} {\langle E_{\rm X} \rangle}. 
\end{equation}

{ Here $L_{e, p}$ ($L_{\rm X}$) is the total power in electrons/protons} (X-rays) per microquasar, $t_{\rm CR}$ ($t_{\rm X}$) is the  CRs (X-rays) source lifespan,  and  $\langle E_{e,p}\rangle$ ($\langle E_{\rm X}\rangle$) is electron/proton (X-ray)  mean energy. The total power in x-rays, $L_{\rm X}$, is a fraction $f_{\rm X}$ of Eddington's luminosity $L_{\rm Edd}$ (e.g., Poutanen, Krolik \& Ryde 1997; Fender, Belloni \& Gallo 2004; Bosch-Ramon, Romero \& Paredes 2006; Romero \& Vila 2008). If the CRs are accelerated at the shocks of the jet of the MQs,  $L_{\rm CR}$ is a fraction $f_{\rm CR}$ of $L_{\rm Edd}$ (because the jet's power is in turn a fraction of $L_{\rm Edd}$). The power in CRs then is $L_{\rm CR} = L_{e} + L_{p} = L_{e} + a\,L_{e} = L_{e} (1+a) = f_{\rm CR}\,L_{\rm Edd}$, where $a$ is the electron/proton power ratio. 

Then, $L_{e}  = f_{\rm CR}\,L_{\rm Edd}/(1+a)$ and $L_{p}  = a\,f_{\rm CR}\,L_{\rm Edd}/(1+a)$, so

\begin{equation}\label{chi2}
\chi = \left(\frac{N_{{\rm io},e}}{\langle E_{e} \rangle} + a\frac{N_{{\rm io},p}}{\langle E_{p} \rangle}\right)\frac{1}{(1+a)}\frac{\langle E_{\rm X} \rangle}{N_{\rm io,X}}\, \frac{f_{\rm CR}\,L_{\rm Edd}}{f_{\rm X}\,L_{\rm Edd}}\,\frac{t_{\rm CR}}{t_{\rm X}}\, \frac{\epsilon_{\rm CR}}{\epsilon_{\rm X}}\,\kappa.
\end{equation}

The active stage of the sources can be assumed to be of the same order because jet launching and X-ray emission  are both intimately related with accretion, so ${t_{\rm CR}}/{t_{\rm X}} \sim 1$. For the same reason we can consider $f_{\rm CR}$ $\sim$ $f_{\rm X}$. 

The power-law shape of the electron/proton distribution implies a great fraction of electrons/protons of energy of a few $m_{e,p}\,c^2$. We {have assumed} 1 keV for $\langle E_{\rm X}\rangle$, 1 MeV for $\langle E_{e}\rangle $ and 1 GeV $\langle E_{p}\rangle $.  For these values, the rate $\langle E_{\rm X} \rangle $/$\langle E_{e} \rangle \sim 10^{-3}$ and $\langle E_{\rm X} \rangle $/$\langle E_{p} \rangle \sim10^{-6}$. Equation~(\ref{chi2}) then reads, 

\begin{equation}\label{chi3}
\chi = \frac{10^{-3}}{(1+a)} \left(\frac{N_{{\rm io},e}}{N_{\rm io,X}} + a\,10^{-3}\,\frac{ N_{{\rm io},p}}{N_{\rm io,X}}\right)\, \frac{\epsilon_{\rm CR}}{\epsilon_{\rm X}}\,\kappa.
\end{equation}

According to our calculations from the previous section,  $N_{\rm io,p}=I^{p}_{\rm 1\,GeV}\sim 10^5$, { and $N_{{\rm io},e}=I^{e}_{\rm 1\,MeV}\sim 1.5\times10^4$} (see Eq.~\ref{eq:1GeVioniz}) and $N_{\rm io, X}\sim 25$,  at $E_{\rm X} = 1$~keV (e.g., Shull \& van Steenberg 1985); then $N_{{\rm io},p}/N_{\rm io,X} \sim 4\times 10^3$ { and  $N_{{\rm io},e}/N_{\rm io,X} \sim 6\times 10^2$.} 

Before reaching the IGM, X-rays photons must traverse the dense cloud in which the system was formed. The optical depth is $\tau = \sigma_{\rm H}\,N_{\rm HI}$, where $\sigma_{\rm H}$ is the bound-free absorption cross-section for H, and $N_{\rm HI}$ is the column density of HI in the primordial clouds. The cross section is $\sigma_{\rm H}(E_{\rm ph}) = 6\times 10^{-18}(13.6\,{\rm eV}/E_{\rm ph})^3$. For $E_{\rm ph} = 1$~keV, and a column density $N_{\rm HI} = 10^{21}$~cm$^{-2}$ (e.g., Wise \& Abel 2012), $\tau$ $\sim$ 0.015. Then, the optical depth results in an attenuation factor of $\epsilon_{\rm X}=e^{-\tau} \sim 0.98$ for X-rays. On the other hand, the jets inject the relativistic particles directly into the IGM and $\epsilon_{\rm CR}=1$. With the above estimates, $\chi$ yields

\begin{equation}\label{chi4}
\chi = \frac{6\times 10^{-3}}{(1+a)} \left( 10^{2} + a\frac{2}{3}\right)\,\kappa.
\end{equation}


In the primordial universe $\kappa$ is expected to be $\sim$ 1. This is because at high redshifts and low  metallicities no colliding wind binaries are expected to exist and all X-ray emission is due to accretion. The jet-disk symbiosis (e.g., Falcke \& Biermann 1995) implies that most X-ray binaries should actually be MQs.

The actual content of the jets of MQs is unknown. Both purely leptonic and lepto-hadronic models have been proposed in the literature (e.g. Bosch-Ramon et al. 2006, Vila et al. 2012). In the case only electrons are accelerated, $a = 0$, $\chi$ is $\sim$ 0.6, and the ionization produced by CRs on the IGM is 60\% of the X-ray ionization.

If $a = 100$, as observed locally in CRs (Ginzburg \& Syrovatskii 1964), the number of ionizations produced by CRs is $\sim$ 0.01 of the  number of ionizations produced by X-rays. 

As these particles are injected directly into the IGM, the slowest diffusion speed is given by Bohm diffusion, one scattering per gyroradius

\begin{equation}
r_{\rm{L}} = 1.1 \mathrm{Mpc} \left(\frac{E}{\mathrm{GeV}}\right)\left(\frac{10^{-18}\mathrm{G}}{B}\right)
\end{equation}

The distance these particles diffuse is $s_{\rm Bohm}=\sqrt{ctr_{\rm L}}$. In the $10^9$yr duration of the reionization period, this gives 5.8 Mpc for 1~GeV protons and 0.18 Mpc for 1 MeV electrons, showing that for a magnetic field $B$=$10^{17} \rm G$ CRs and hard X-rays can propagate on comparable distances.

\section{Discussion}\label{disc}

The first generation of MQs started the process of reionization of the IGM a few Myr after the formation of the first stars. As soon as the firsts Pop III stars imploded creating the first stellar mass black holes, MQ jets were switched on. We then arrive at a picture where early reionization was achieved locally by UV from stars and by X-ray ray photons from accreting black holes at medium scales but still inside the primordial gas clouds. The jets associated with the accretion processes transport energy and create CR sources at their termination points, outside the clouds. Then, the reionization of the IGM might have started at a very early epoch. The cooling time of low-energy CRs, which constitute the bulk of the primordial CR population, is such as to allow these  particles to reach epochs with $z\sim 5$. There is then an accumulative effect through the whole reionization era. At some point, CR contributions from jets of high-redshift AGNs will start to play a role as well, along with the X-ray photons from putative intermediate mass black holes and the accretion disks of early quasars. Regarding CR production in Pop III supernovae (e.g., Stacy \& Bromm 2007), they will be mostly confined to the original clouds surrounding the stars. It is not even clear whether such CRs would be produced at all, since very massive stars might implode directly into black holes without going through an explosion (e.g., Mirabel \& Rodrigues 2003). The final fate of Pop III stars might be long gamma-ray bursts (e.g.,  M\'esz\'aros \& Rees 2010), which might inject additional CRs into the IGM, enhancing the effects discussed in this Letter. 

The reionization of the universe was likely a complex process, involving different types of mechanisms operating simultaneously with effects on different scales.  Both high-energy photons and charged relativistic particles seem to have played a role in this crucial period of the cosmic history.

\begin{acknowledgements}
We thank Alejandra Kandus for insightful discussions on primordial magnetic fields and an anonymous referee for constructive comments. This work was supported by  PICT 2012-00878, Pr\'estamo BID (ANPCyT) and grant AYA 2013-47337-C3-1-P (MINECO, Spain). 
\end{acknowledgements}



\begin{thebibliography}{20}

  


\bibitem{AIRES} Aires Manual, S. Sciutto. $www2.fisica.unlp.edu.ar/auger\/aires/eg\_Aires.html$
\bibitem{Nath93} Biermann, P. \& Nath, B. 1993 \mnras, 265,241
\bibitem{Belikov09} Belikov, A. \& Hooper, D., Phys.Rev. D, 80, 3, 035007
\bibitem{bordas09} Bordas, P., Bosch-Ramon, V., Paredes, J.~M., Perucho, M. 2009, A\& A, 497, 325
\bibitem{bosch-ramon06}Bosch-Ramon, V., Romero, G.~E. \& Paredes, J.~M. 2006, A \& A, 447,263
\bibitem{bromm13}Bromm, V. 2013, Rep. Progr. Phys., 76, 112901
\bibitem{Cumberbatch10} Cumberbatch et al., 2010, Phys. Rev. D, 82, 103508 
\bibitem{drury83}Drury, L. O. 1983, Rep. Progr. Phys., 46, 973
\bibitem{ellis12} Ellis G., Maartens R. \& MacCallum M. 2012, Relativistic Cosmology, Cambridge University Press, Cambridge
\bibitem{falcke95}Falcke, H. \&  Biermann, P.~L. 1995, \aa, 293, 665
\bibitem{fender04}Fender, R.~P., Belloni, T.~M. \& Gallo, E. 2004, \mnras, 355, 1105
\bibitem{ginzburg64}Ginzburg, V.L. \& Syrovatskii, S.I. 1964, The Origin of Cosmic Rays, Pergamon Press, Oxford
\bibitem{heinz02}Heinz, S., Sunyaev, R. 2002, A\&A, 390, 751
\bibitem{iocco10} Iocco, F. AIP Conf. Proc, 1241, pp. 379-387 (2010).
\bibitem{kogut03}Kogut, A. et al. 2003, ApJS 148,161
\bibitem{krticka06} Krti\v{c}ka, J. \& Kub\'at, J. 2006, A \& A, 446, 1039
\bibitem{loeb10}Loeb, A. 2010, How Did the First Stars and Galaxies Form?, Princeton University Press, Princeton
\bibitem{meszaros}M\'esz\'aros, P. \& Rees, M.~J. 2010, \apj, 715, 967
\bibitem{middleton13}Middleton, M.~J.,  Miller-Jones, J.~C.~A., Markoff, S., Fender, R. et al. 2013,
\nat, 493, 187
\bibitem{mirabel99}Mirabel, I.~F. \& Rodr\'iguez, L.~F. 1999, Annu. Rev. Astron. Astrophys., 37, 409
\bibitem{mirabel03}Mirabel, I.~F. \& Rodrigues, I. 2003, Science, 300, 1119
\bibitem{mirabel11}Mirabel, I.~F., Dijkstra, M.,  Laurent, P., Loeb, A. \& Pritchard, J.~R. 2011, A\& A 528, A149
\bibitem{Natarajan09} Natarajan, A. \& Schwarz, D.,2010, Phys. Rev. D, 81, 12, 123510
\bibitem{pakull10}Pakull, M.~W., Soria, R., Motch, C. 2010, \nat, 466, 209
\bibitem{Planck} Planck collaboration, arXiv:1303.5076
\bibitem{poutanem97}Poutanem, J., Krolik, J.~H., \& Ryde, F. 1997, \mnras, 292, L21
\bibitem{romero08}Romero, G.~E., \& Vila, G.~S. 2008, A\& A, 485, 623
\bibitem{shull85}Shull, J.~M., \& van Steenberg, M.~E. 1985, \apj, 298, 268
\bibitem{Spitzer68}Spitzer, L. \& Tomasko, M., 1968, \apj, 152, 971 
\bibitem{stacy07}Stacy, A. \& Bromm, V. 2007, \mnras, 382, 229
\bibitem{Valdes10} Valde\'s, M. et al. 2010, \mnras, 404, 1569
\bibitem{vila10}Vila, G.~S. \& Romero, G.~E. 2010, \mnras, 403, 1457
\bibitem{vila12}Vila, G.~S., Romero, G.~E., \& Casco, N.~A. 2012, A\& A, 538, A97
\bibitem{Wilow12} Widrow, L et al. 2012, Space Sc. Rev., 166,4,37
\bibitem{Wise10} Wise, J. \& Abel, T. 2010, \apj, 684, 1
\bibitem{zealey80}Zealey, W.~J., Dopita, M.~A. \& Malin, D.~F., \mnras, 192, 731

\end{thebibliography}
\end{document}